\documentclass[draft]{agujournal2019}
\usepackage{url} %
\usepackage{lineno}
\usepackage[inline]{trackchanges} %
\usepackage{listings}

\usepackage{soul}
\usepackage{amsmath, amsfonts, amssymb}
\usepackage{cleveref}
\usepackage{subcaption}

\renewcommand{\vec}[1]{\mathbf{#1}}

\draftfalse

\journalname{journal}

\graphicspath{{./figures/}}

\begin{document}

\title{Accelerating large-eddy simulations of clouds with Tensor Processing Units}

\authors{Sheide Chammas\affil{1}, Qing Wang\affil{1}, Tapio Schneider\affil{1,2}, Matthias Ihme\affil{1,3}, Yi-fan Chen\affil{1}, and John Anderson\affil{1}}

\affiliation{1}{Google LLC}
\affiliation{2}{California Institute of Technology}
\affiliation{3}{Stanford University}

\correspondingauthor{Sheide Chammas}{sheide@google.com}

\begin{keypoints} %
\item We introduce a large-eddy simulation (LES) framework that runs on Tensor Processing Units (TPUs, accelerators designed for machine learning) 
\item The fidelity of the LES is established by reproducing aircraft observations of nocturnal stratocumulus clouds over the Pacific
\item The LES exhibit unprecedented scalability on TPUs, enabling the large-scale generation of training data for cloud parameterizations
\end{keypoints}

\begin{abstract}
Clouds, especially low clouds, are crucial for regulating Earth's energy balance and mediating the response of the climate system to changes in greenhouse gas concentrations. Despite their importance for climate, they remain relatively poorly understood and are inaccurately represented in climate models. A principal reason is that the high computational expense of simulating them with large-eddy simulations (LES) has inhibited broad and systematic numerical experimentation and the generation of large datasets for training parametrization schemes for climate models. Here we demonstrate LES of low clouds on Tensor Processing Units (TPUs), application-specific integrated circuits that were originally developed for machine learning applications. We show that TPUs in conjunction with tailored software implementations can be used to simulate computationally challenging stratocumulus clouds in conditions observed during the Dynamics and Chemistry of Marine Stratocumulus (DYCOMS) field study. The TPU-based LES code successfully reproduces clouds during DYCOMS and opens up the large computational resources available on TPUs to cloud simulations. The code enables unprecedented weak and strong scaling of LES, making it possible, for example, to simulate stratocumulus with $10\times$ speedup over real-time evolution in domains with a $34.7~\mathrm{km} \times 53.8~\mathrm{km}$ horizontal cross section. The results open up new avenues for computational experiments and for substantially enlarging the sample of LES available to train parameterizations of low clouds.
\end{abstract}

\section*{Plain Language Summary}
The study of clouds has been impeded by, among other factors, limitations in our ability to simulate them rapidly and on sufficiently large domains. In particular, computational limitations in simulating low clouds are among the reasons for the difficulties of representing them accurately in climate models; this is one of the dominant uncertainties in climate predictions. This paper demonstrates how the large computing power available on Tensor Processing Units (integrated circuits originally designed for machine learning applications) can be harnessed for simulating low clouds. We demonstrate the largest simulations of low clouds to date, with hundreds of billions of variables, and we document their fidelity to aircraft observations. The results open up the large computational resources available on TPUs, hitherto primarily used for machine learning, to the study of clouds in the climate system.

\section{Introduction}

Scientific progress accelerates when it is possible to cycle rapidly through the knowledge discovery loop: design and conduct experiments, learn from the experiments, and design and conduct new experiments to test and refine models and hypotheses with the information obtained from them \cite{NASEM22a}. In the computational sciences, experiments are conducted numerically, and the ability to cycle through the knowledge discovery loop has advanced hand-in-hand with the evolution of computer hardware. The atmospheric sciences represent a prime example of advances in computer hardware enabling and accelerating scientific progress. The first experiments with two-dimensional atmosphere models \cite{Charney50o} and, soon thereafter, with quasigeostrophic two-layer models \cite{Phillips54,Phillips56} only allowed simulations that were slower than or comparable with the real-time evolution of the atmosphere. The first experiments  using general circulation models similarly pushed the envelope of what was computationally feasible at the time \cite{Smagorinsky63,Smagorinsky65,Manabe65}. Once such simulations of atmospheric flows, albeit at coarse resolution, became routine and rapidly executable, systematic exploration and experimentation followed, enabling rapid progress in our understanding of the atmosphere's general circulation, from its dependence on planetary characteristics such as planetary radius and rotation rate \cite{Williams88a,Williams88b}, over the nature of atmospheric turbulence \cite{Rhines75,Rhines79a,Held96,Held99b,Schneider06a,Schneider06b}, to  elucidating  the hydrologic cycle \cite{Manabe75a,Rind92a,Held06,Allen02,Chou04,OGorman08b,Schneider10a}. Similarly, our understanding of deep convective clouds advanced substantially once deep-convection resolving simulations in limited areas became routinely feasible \cite{Held93b,Tompkins98a,Cronin14b,Cronin15a,Wing18a}. In contrast, our understanding of the dynamics of low clouds is in its infancy. We do not have quantitative theories of their response to climate change \cite{Bretherton15a}, and shortcomings in their representation in climate models have long dominated uncertainties in climate projections \cite{Cess90a,Cess96a,Bony05a,Dufresne08a,Vial13a,Brient16b,Brient16a,Webb06a,Webb13b,Zelinka17a}. Numerical experiments have been limited to studies that have explored a few dozen canonical situations, mostly in the tropics \cite{Siebesma03,Stevens05a,Rauber07a,Caldwell09a,Sandu11a,Zhang12b,Zhang13a,Blossey13a,Blossey16a,Schalkwijk15a,Tan16b,Tan17a}. Broader exploration has been limited by the computational expense necessary to resolve the meter-scale dynamics of low clouds in large-eddy simulations (LES).

Here we take the next step in the co-evolution of science and computing hardware by demonstrating LES of low clouds on Tensor Processing Units (TPUs). TPUs are application-specific integrated circuits (ASICs),  originally developed for machine learning applications, which are dominated by dense vector and matrix computations \cite{Jouppi17a}. The current TPU  architecture integrates 4,096 chips into a so-called TPU Pod, which achieves 1.1 exaflops in aggregate at half precision. TPUs are publicly available for cloud computing and can be leveraged for fluids simulations \cite{Wang2022-ln} and other scientific computing tasks~\cite{BELLETTI_ETAL_ARXIV2019,LU_CHEN_HECHTMAN_WANG_ANDERSON_ARXIV2020,PEDERSON_ETAL_JCTC2022}, with remarkable computational throughput and scalability. Large, high-bandwidth memory and fast chip-to-chip interconnects (currently 1.1 PB/s) contribute to the performance of TPUs and alleviate bottlenecks that computational fluid dynamics (CFD) applications typically face on accelerator platforms \cite{Balaji21a}. However, the native half- or single-precision arithmetic of TPUs can also create challenges in CFD applications~\cite{Wang2022-ln}.

The objective of this study is to evaluate the throughput and scalability achievable on TPUs in simulations of subtropical stratocumulus clouds under conditions encountered during the Dynamics and Chemistry of Marine Stratocumulus (DYCOMS) field study \cite{Stevens05a}. Stratocumulus clouds are a particularly good testbed for low-cloud simulations for two reasons: First, they are the most frequent cloud type on Earth, covering about 20\% of tropical oceans, with an outsize impact on Earth's energy balance \cite{Wood12a}. Reductions or increases in the area they cover by a mere 4\% can have an impact on Earth's surface temperature comparable to doubling or halving atmospheric carbon dioxide concentrations \cite{Randall84c}. Second, they are notoriously difficult to simulate, even in LES, because key processes responsible for their maintenance, such as turbulent entrainment of air across the often sharp temperature inversions at their tops, occur on scales of meters \cite{Mellado16a}. The resulting numerical challenges lead to large differences among various LES codes owing to differences in the numerical discretizations \cite{Stevens05a,Pressel17a}. For example, weighted essentially non-oscillatory (WENO) advection schemes at resolutions of $O(10~\mathrm{m})$ lead to more faithful simulations---relative to field measurements---than centered difference advection schemes at resolutions of $O(1~\mathrm{m})$ \cite[their supplementary Fig.~3]{Schneider19a}. These two reasons make progress in simulating subtropical stratocumulus both important and challenging.

This paper is structured as follows. Section~\ref{s:model} describes the governing equations, numerical methods, and TPU-specific implementation decisions in our LES code. Section~\ref{s:validation} presents a dry buoyant bubble and a density current as validation examples of the code. Section~\ref{s:DYCOMS} presents the DYCOMS simulations, including comparisons with field data and a scaling analysis of the simulations. Section~\ref{s:concl} summarizes the conclusions and new opportunities afforded by this TPU-enabled  cloud-simulation capability.

\section{Model Formulation, Numerics, and TPU implementation}\label{s:model}

\subsection{Governing Equations}

Our LES simulates the anelastic equations for moist air, understood to be an ideal admixture of dry air, water vapor, and any condensed water that is suspended in and moves with the air. Precipitating condensate (e.g., rain and snow) is not considered part of the working fluid, and the suspended constituents of the moist air are taken to be in local thermodynamic equilibrium. By Gibbs' phase rule, then, a complete thermodynamic description of this system with two components (dry air and water) and three phases (water vapor, liquid water, ice) requires specification of two thermodynamic variables, in addition to the density $\rho$ and pressure $p$ of the moist air. We choose the total water specific humidity $q_t$ (total mass of water per unit mass of moist air) and liquid-ice potential temperature $\theta_l$ \cite{Tripoli81a}. This choice of thermodynamic variables is advantageous because both the total specific humidity $q_t$ and (approximately) the liquid-ice potential temperature $\theta_l$ are materially conserved even in the presence of reversible phase transitions of water. The temperature $T$ and specific humidities $q_l$ and $q_i$ of cloud liquid and ice can be computed from the other thermodynamic variables.

The anelastic approximation eliminates physically insignificant acoustic waves by linearizing the density $\rho(x,y,z,t) = \rho_0(z) + \rho'(x,y,z,t) $ and pressure $p(x,y,z,t) = p_0(z) + p'(x,y,z,t)$ around a dry reference state with density $\rho_0(z)$ and hydrostatic pressure $p_0(z)$, which depend only on altitude $z$. Here, reference state variables are indicated by a subscript $0$, and perturbation variables by primes. Perturbation variables are retained only where they affect accelerations. The reference density and pressure depend only on the vertical coordinate $z$ and are in hydrostatic balance,
\begin{equation}
    \frac{\partial p_0(z)}{\partial z} = -\rho_0(z) g.
    \label{eq:hydro-static}
\end{equation}
For energetic consistency, the reference state needs to be adiabatic, i.e., the reference potential temperature $\theta_0$ needs to be constant \cite{Bannon96a,Pauluis08a}. Therefore, 
\begin{align}
     T_0 &= \theta_0 \left(\frac{p_0}{p_{00}}\right)^{R_d/c_{pd}}, \\
     p_0 &= p_{00}\left(1-\frac{gz}{c_{pd}\theta_0}\right)^{c_{pd}/R_d}, \\
     \rho_0 & = \frac{p_0}{R_d T_0}.
\end{align}
Table~\ref{t:constants} summarizes the thermodynamic constants and other parameters used in the present study.
\begin{table}[!htb!]
 \caption{Thermodynamic constants and other parameters used in this study.}\label{t:constants}
 \centering
 \begin{tabular}{c l l}
 \hline
  Symbol  & \multicolumn{1}{c}{Name} & \multicolumn{1}{c}{Value}  \\
 \hline
    $p_{00}$                & Constant reference pressure & $1000~\mathrm{hPa}$ \\
    $\theta_0$              & Reference potential temperature & $290~\mathrm{K}$ \\
    $R_d$                   & Gas constant of dry air & $287~\mathrm{J~(kg~K)^{-1}}$\\
    $R_v$                   & Gas constant of water vapor & $461.89~\mathrm{J~(kg~K)^{-1}}$\\
    $c_{pd}$                & Isobaric specific heat capacity of dry air & $1004.5~\mathrm{J~(kg~K)^{-1}}$ \\
    $c_{pv}$                & Isobaric specific heat capacity of water vapor & $1859.5~\mathrm{J~(kg~K)^{-1}}$ \\
    $c_{l}$                & Specific heat capacity of liquid water & $4181~\mathrm{J~(kg~K)^{-1}}$ \\
    $c_{i}$                & Specific heat capacity of ice & $2100~\mathrm{J~(kg~K)^{-1}}$ \\
    $L_{v,0}$               & Specific latent heat of vaporization & $2.47~\mathrm{MJ~kg^{-1}}$\\
    $L_{s,0}$               & Specific latent heat of sublimation & $2.83~\mathrm{MJ~kg^{-1}}$ \\
    $T_f$                   & Freezing point temperature
    & $273.15~\mathrm{K}$\\
    $f$                     & Coriolis parameter & $7.62 \times 10^{-5}~\mathrm{s^{-1}}$\\
    $g$                     & Gravitational acceleration & $9.81~\mathrm{m~s^{-2}}$ \\
    $c_s$                   & Smagorinsky constant & $0.18$ \\
    $\mathrm{Pr}$           & Turbulent Prandtl number & $0.4$ \\
    $\mathrm{Sc}_{q_t}$     & Turbulent Schmidt number of water & $0.4$\\
 \hline
 \end{tabular}
 \end{table}

Thermodynamic consistency of the anelastic system requires that thermodynamic quantities are evaluated with the reference pressure $p_0(z)$ \cite{Pauluis08a}. Therefore, the liquid-ice potential temperature we use is
\begin{equation}
\theta_l(T, q_l, q_i; p_0) = \frac{T}{\Pi}\left( 1 - \frac{L_{v,0} q_l + L_{s,0} q_i}{c_{pm} T} \right),
\end{equation}
where
\begin{equation}
    \Pi = \left( \frac{p_0(z)}{p_{00}} \right)^{R_m/c_{pm}}
\end{equation}
is the Exner function, evaluated with the altitude-dependent reference pressure $p_0(z)$ and the constant pressure $p_{00}$. We take water vapor and suspended cloud condensate into account in the gas ``constant``  $R_m = (1-q_t)R_d + (q_t-q_c)R_v$ (which is not constant because it depends on the total specific humidity $q_t$ and condensate specific humidity $q_c=q_l + q_i$) and in the isobaric specific heat $c_{pm} = (1-q_t)c_{pd} + (q_t-q_c)c_{pv} + q_l c_l + q_i c_i$. 

With these definitions, the anelastic governing equations in conservation form are
\begin{align}
\nabla\cdot (\rho_0\vec{u}) & = 0,\label{e:conti}\\
\frac{\partial (\rho_0 \vec{u})}{\partial t} + \nabla\cdot \left( \rho_0 \vec{u} \otimes \vec{u}\right) &= - \rho_0 \nabla \left( \alpha_0  p' \right)
+ \rho_0 b \vec{k} - f\vec{k} \times \rho_0 (\vec{u}-\vec{u}_g) + \nabla\cdot (\rho_0 \mathbf{\sigma}),\label{e:momentum}\\
 \frac{\partial (\rho_0 \theta_l)}{\partial t} + \nabla\cdot \left(\rho_0 \vec{u} \theta_l \right) 
 & = -\frac{1}{c_{pm} \Pi} \nabla\cdot (\rho_0 \vec{F}_R) + \rho_0 w_{\mathrm{sub}} \frac{\partial \theta_l}{\partial z} + \frac{1}{\mathrm{Pr}} \nabla\cdot \bigl(\rho_0 \nu_t \nabla \theta_l \bigr), \label{e:potential_temperature} \\ %
\frac{\partial (\rho_0 q_t)}{\partial t}+ \nabla\cdot (\rho_0 \vec{u} q_t)
& = \rho_0 w_{\mathrm{sub}} \frac{\partial q_t}{\partial z} + \frac{1}{\mathrm{Sc}_{q_t}} \nabla\cdot (\rho_0 \nu_t \nabla q_t). \label{e:total_humidity}   
\end{align}
Here,
\begin{equation}
    b = g \frac{\alpha(\theta_l, q_t, p_0) - \alpha_0(z)}{\alpha_0(z)}
\end{equation}
is the buoyancy, and $\alpha_0 = 1/\rho_0$ and $\alpha = 1/\rho$ are specific volumes. The specific volume $\alpha(\theta_l, q_t, p_0)$ is calculated from the approximate equation of state, again with the reference pressure $p_0$ in place of the total pressure,
\[
\alpha = \frac{R_m T}{p_0}.
\]
Neglected in these equations is differential settling of condensate relative to the surrounding air and all precipitation processes. %
Table~\ref{t:vars} lists the variables we use.
\begin{table}[!htb!]
 \caption{Definitions of Variables}\label{t:vars}
 \centering
 \begin{tabular}{c l l}
 \hline
  Variable  & \multicolumn{1}{c}{Definition} & \multicolumn{1}{c}{Units}  \\
 \hline
   $\rho$                   & Density of moist air          & $\mathrm{kg~m^{-3}}$\\ 
   $\alpha$                 & Specific volume of moist air  &
   $\mathrm{m^3~kg^{-1}}$ \\
   $\vec{u}$                & Velocity of moist air         &
   $\mathrm{m~s^{-1}}$\\
   $\vec{u}_g$              & Prescribed geostrophic velocity  &
   $\mathrm{m~s^{-1}}$\\
   $w_{\mathrm{sub}}$       & Prescribed subsidence velocity
   & $\mathrm{m~s^{-1}}$ \\
   $p$                      & Pressure                      & $\mathrm{Pa}$\\
   $b$                      & Buoyancy                      &
   $\mathrm{m~s^{-2}}$ \\
    $\vec{k}$               & Vertical unit vector \\
   $T$                      & Temperature                   & $\mathrm{K}$ \\
   $R_m$                    & Specific gas ``constant'' of moist air & $\mathrm{J~kg^{-1}~K^{-1}}$ \\
    $c_{pm}$                & Isobaric specific heat of moist air & $\mathrm{J~kg^{-1}~K^{-1}}$ \\
   $\mathbf{\sigma}$           & Subgrid-scale stress per unit mass      & $\mathrm{m^2~s^{-2}}$ \\
   $\vec{F}_R$              & Radiative energy flux         & $\mathrm{W~m~kg^{-1}}$ \\
   $q_t$                  & Total water specific humidity & $\mathrm{kg/kg}$\\
  $q_v$                     & Water vapor specific humidity & $\mathrm{kg/kg}$\\
  $q_l$                     & Liquid water specific humidity & $\mathrm{kg/kg}$\\
  $q_i$                     & Ice specific humidity         & $\mathrm{kg/kg}$\\
 $\nu_t$                    & Turbulent viscosity
    & $\mathrm{m^2~s^{-1}}$ \\
    $z$                     & Altitude
    & $\mathrm{m}$\\
 \hline
 \end{tabular}
 \end{table}
 The perturbation pressure $p'$ is obtained as solution to a Poisson equation, which follows by taking the divergence of the momentum equation. The numerical algorithm for solving~\cref{e:conti,e:momentum,e:potential_temperature,e:total_humidity} is discussed in section~\ref{s:numerics}.
 
 \subsection{Saturation Adjustment}

The temperature $T$ and the partitioning of total water mass into the liquid phase (specific humidity $q_l$) and ice phase (specific humidity $q_i$) are obtained from $\theta_l$, $q_t$, and the reference pressure $p_0$ by a saturation adjustment procedure \cite{Tao89a}. This amounts to solving 
\begin{equation}\label{e:sat_adj}
    \theta_l^* - \theta_l = 0,
\end{equation}
where $\theta_l^*(T; p_0) = \theta_l(T, q_l^*, q_i^*; p_0)$ is the liquid-ice potential temperature at saturation, that is, with 
\begin{equation}\label{e:liquid}
    q_l^* = \max\bigl[0, q_t - q_v^*(T, p_0)\bigr] \mathcal{H}(T - T_f)
\end{equation}
and
\begin{equation}\label{e:ice}
    q_i^* = \max\bigl[0, q_t - q_v^*(T, p_0)\bigr] \mathcal{H}(T_f - T).
\end{equation}
Here, $q_v^*$ is the saturation specific humidity, calculated as in \citeA{Sridhar22a}, $\mathcal{H}$ is the Heaviside step function, and $T_f$ is the freezing point temperature. We solve the resulting nonlinear problem \eqref{e:sat_adj} with the secant method. In the presence of mixed-phase clouds, the requirement of instantaneous thermodynamic equilibrium should be relaxed, for example, by replacing the Heaviside function in Eqs.~\eqref{e:liquid} and \eqref{e:ice} by a continuous phase partitioning function \cite<e.g.,>{Tao89a,Pressel15a}, or by carrying separate prognostic variables for liquid and ice specific humidities. However, in the examples here we focus on warm clouds with only liquid.  

\subsection{Subgrid-scale Models}

We model subgrid-scale fluxes with the turbulent viscosity model of \citeA{Lilly62a} and \citeA{Smagorinsky63}. In this model, the turbulent viscosity is represented as 
\begin{equation}
    \nu_t = (c_s \Delta)^2 f_B S.
\end{equation}
where $S = \| \mathbf{S} \|_2$ is the 2-norm of the strain rate tensor $\mathbf{S} =0.5 \bigl[\nabla \vec{u} + (\nabla \vec{u})^T \bigr]$ for the resolved velocities $\vec{u}$; $c_s$ is the Smagorinsky constant (Table~\ref{t:constants}); and $\Delta = (\Delta x \Delta y \Delta z)^{1/3}$ is the geometric mean of the grid spacings in the three space directions. The buoyancy factor $0 \le f_B \le 1$ limits the mixing length in the vertical in the case of stable stratification; it is computed from the moist buoyancy frequency \cite{Durran82a} as described in \citeA{Pressel17a}. The diffusivities of the liquid-ice potential temperature and total specific humidity are obtained from the turbulent viscosity $\nu_t$ by division by constant turbulent Prandtl and Schmidt numbers (Table~\ref{t:constants}).

To emulate a radiation condition at the upper boundary, we include a sponge layer that occupies the top 5\% of the domain and absorbs upward propagating waves. The sponge is implemented as a linear Rayleigh damping layer \cite{Durran83a}, in which the horizontal velocity is  relaxed toward the geostrophic wind velocity and the vertical velocity is relaxed to zero. To avoid reflections at the interface between the sponge layer and the undamped flow outside, we use a relaxation coefficient that ensures a gradual onset of the sponge layer \cite{Klemp1978-lp}, reaching $0.25~\mathrm{s^{-1}}$ at the top of the domain.

\subsection{Numerical Solution}\label{s:numerics}

We discretize the governing equations with the finite-difference method. All discrete operators are expressed on a collocated mesh. All diffusion terms are computed with a 2nd-order central difference scheme. The advection terms in \cref{e:momentum,e:potential_temperature,e:total_humidity} are discretized with the 3rd-order QUICK (Quadratic Upstream Interpolation for Convective Kinematics) scheme.  While the QUICK scheme is upwind-biased, it is not monotonicity preserving. This implies that in interaction with subgrid-scale diffusion, it can create spurious mixing in regions of sharp gradients, for example, at a sharp inversion topping a boundary layer, with potentially deleterious effects on the simulation of stratocumulus clouds \cite{Bretherton99a,Pressel17a}. As we will see, these effects are real but minor in our case. Alternatively, one may construct a monotone version of the QUICK scheme by applying flux limiters \cite{Zalesak79a,Stevens96a}. To test the effects of the advection scheme near the inversion, we also implemented a 3rd-order WENO scheme, reconstructing the advective fluxes on cell faces and using a Lax-Friedrichs numerical flux, as in standard finite-volume methods.

An explicit iterative scheme~\cite{Wang2022-ln} is employed for the time advancement of the numerical solutions. This scheme provides an iterative representation to the Crank-Nicolson method, which avoids the computational complexity of solving a high-dimensional linear system of equations. Specifically, the momentum equation \eqref{e:momentum} is solved with a predictor-corrector approach. At the prediction step of sub-iteration $k+1$, the momentum equation is solved in discrete form as
\begin{align}
    \frac{\widehat{\rho_0 \vec{u}} - (\rho_0 \vec{u})^n}{\Delta t}
    &= -\rho_0\nabla \left(\alpha_0{p^\prime}^k\right) + \vec{R}^{n+\frac{1}{2}},
    \label{eq:momentum_prediction_step}
\end{align}
with
\begin{equation}
    \vec{R}^{n+\frac{1}{2}}= -\nabla\cdot[(\rho_0 \vec{u})^{n+\frac{1}{2}}\otimes\vec{u}^{n+\frac{1}{2}}] + \nabla\cdot\left(\rho_0\sigma^{n+\frac{1}{2}}\right)  + \rho_0 b^{n+\frac{1}{2}}\vec{k} -f\vec{k}\times\rho_0\left(\vec{u}^{n+\frac{1}{2}}-\vec{u}_g\right),
\end{equation}
where $\widehat{(\cdot)}$ denotes a prediction of a variable at step $n+1$ in sub-iteration $k+1$; $(\cdot)^k$ is the solution of a variable at step $n+1$ obtained from sub-iteration $k$. Variables at state $(\cdot)^{n+\frac{1}{2}}$ are estimated as $(\cdot)^{n+\frac{1}{2}}=[(\cdot)^{k}+(\cdot)^{n}]/2$. Note that the prediction of the momentum $\widehat{\rho_0\vec{u}}$ in sub-iteration $k+1$ is evaluated with the pressure from the previous sub-iteration. The correct momentum $(\rho_0\vec{u})^{k+1}$ needs to be computed with the pressure at sub-iteration $k+1$, which can be expressed similarly to~\cref{eq:momentum_prediction_step} as
\begin{equation}
    \frac{(\rho_0 \vec{u})^{k+1} - (\rho_0 \vec{u})^n}{\Delta t} = -\rho_0\nabla \left(\alpha_0{p^\prime}^{k+1}\right) + \vec{R}^{n+\frac{1}{2}}.
    \label{eq:momentum_k+1}
\end{equation}
Subtracting~\cref{eq:momentum_prediction_step} from~\cref{eq:momentum_k+1} yields
\begin{equation}
    \frac{(\rho_0 \vec{u})^{k+1} - \widehat{\rho_0 \vec{u}}}{\Delta t} = -\rho_0\nabla \left(\alpha_0 {p^\prime}^{k+1} - \alpha_0 {p^\prime}^k\right)=-\rho_0\nabla\left(\alpha_0\delta p\right),
    \label{eq:momentum_correction}
\end{equation}
where $\delta p=p'^{k+1} - p'^k$ is the pressure correction from sub-iteration $k$ to $k+1$. 

Taking the divergence of~\cref{eq:momentum_correction} and applying mass conservation at sub-iteration $k+1$ leads to a generalized Poisson equation for the pressure correction:
\begin{align}
    \nabla^2\left(\alpha_0\delta p\right) &= \frac{\alpha_0}{\Delta t}\left[\nabla\cdot(\widehat{\rho_0 \vec{u}}) - \nabla\cdot(\rho_0 \vec{u})^{k+1}\right] \nonumber \\
    &= \frac{\alpha_0}{\Delta t}\nabla\cdot(\widehat{\rho_0 \vec{u}}).
    \label{eq:pressure_correction}
\end{align}
To ensure numerical consistency and eliminate the checkerboard effect due to the collocated mesh representation, we introduce an additional correction term when solving~\cref{eq:pressure_correction}, which is described in~\ref{sec:appendix_a}.

We apply homogeneous Neumann boundary conditions on the pressure, assuming a vanishing correction of the mass flux: $(\rho_0 \vec{u})^{k+1} - \widehat{\rho_0 \vec{u}}=0$. Solving the Poisson equation \eqref{eq:pressure_correction} subject to the boundary conditions provides the pressure correction. We solve the Poisson equation iteratively with the weighted Jacobi method. The momentum and pressure at sub-iteration $k+1$  are then updated as
\begin{align}
    (\rho_0 \vec{u})^{k+1} &= \widehat{\rho_0 \vec{u}} - \Delta t \rho_0\nabla(\alpha_0\delta p), \\
    p^{k+1} &= p^k + \delta p.
\end{align}

The scalar transport equations are discretized with the same numerical scheme as the momentum equation in~\cref{eq:momentum_prediction_step}. For a generic primitive scalar $\phi$, the prediction of its value at step $n+\frac{3}{2}$ from sub-iteration $k+1$ is represented as
\begin{equation}
    \frac{(\rho_0 \phi)^{k+1} - (\rho_0 \phi)^{n+\frac{1}{2}}}{\Delta t} = -\nabla\cdot[(\rho_0 \boldsymbol{u})^{n+1} \phi^{n + 1}] + \nabla\cdot[\rho_0\mathcal{D}_{\phi}^{n + 1}\nabla \phi^{n + 1}] + S_\phi(\phi^{n + 1}, \boldsymbol{u}^{n + 1}),
    \label{eq:scalar_prediction_step}
\end{equation}
where $\mathcal{D}_\phi$ and $S_\phi$ are the diffusivity and multi-physics source term of $\phi$, respectively. Note that the advancement of scalars is a half step ahead of the momentum, which is from step $n+\frac{1}{2}$ to $n + \frac{3}{2}$. This staggered treatment in time advancement improves the convergence of the iterative time-integration scheme~\cite{Pierce2001-nw}. Specifically, terms on the right-hand side of~\cref{eq:scalar_prediction_step} are evaluated at step $n + 1$, where $\boldsymbol{u}^{n + 1} = \boldsymbol{u}^k$ can be obtained from the latest prediction of the velocity at sub-iteration $k$, and $\phi^{n + 1} = (\phi^k + \phi^{n+\frac{1}{2}}) / 2$ is interpolated linearly between its predicted value at sub-iteration $k$ and the solution at the previous step $n + \frac{1}{2}$.

We have verified that the numerical discretization exactly (to machine precision) conserves domain-integrals of scalars in the absence of non-conservative sources and sinks.

\subsection{TPU Implementation\label{subsec:tpu_impl}}

The discrete formulations are implemented in TensorFlow, to support  execution on different computing architectures and integration with machine learning approaches. In the present study, all computations are performed on TPUs; the host CPUs are used for data input and output only. 

At the beginning of each simulation, the simulator code is compiled by the Accelerated Linear Algebra (XLA) compiler with the just-in-time (JIT) approach, which builds a TensorFlow graph. This approach reduces the computational cost at runtime significantly, which is particularly beneficial for simulations with repeated steps. The representations of the three-dimensional data structure and numerical operators are designed to optimize the performance within the TensorFlow programming paradigm~\cite{Wang2022-ln}. The graph is subsequently replicated onto each TPU for computation. The initial flow field data are distributed onto each TPU as input to the distributed graph.

On TPUs, the efficiency of partitioning is anisotropic along different spatial dimensions. This behavior results from the data structures that are designed for optimal computational efficiency. With this programming strategy, partitioning in different directions leads to different TensorFlow graph structures. As a result, partitioning along the first dimension of the allocated 3D tensors is more efficient than along the other two dimensions~\cite{Wang2022-ln}. We investigate the scaling of our simulation framework for different partitions in~\cref{subsec:scaling}, with an assessment of implications for cloud simulations.

\section{Validation Study} 
\label{s:validation}

To validate the numerical scheme and model formulation, we consider two test cases that are widely used for validation  and are relevant to the buoyancy-driven dynamics prevalent in the atmosphere. The first case is a density current consisting of a two-dimensional negatively buoyant dry bubble impinging on a surface \cite{Straka1993-wk}; the second case is a rising buoyant bubble \cite{Bryan04a}. 

\subsection{Density Current}

The density current configuration consists of an initial perturbation to a uniform potential temperature field. The initial perturbation's amplitude peaks at $-15~\mathrm{K}$ and has a horizontal radius of $4~\mathrm{km}$ and a vertical radius of $2~\mathrm{km}$. The two-dimensional domain is $51.2~\mathrm{km}$ wide and $6.4~\mathrm{km}$ high. As in \citeA{Pressel15a}, we use periodic horizontal boundary conditions instead of the no-flux boundary conditions in \citeA{Straka1993-wk}. This benchmark case has an added significance in stratocumulus simulations because the density current’s perturbation amplitude is of the same magnitude as the jump in temperature observed across the entrainment interfacial layer at the cloud top.

\begin{figure}[!htb!]
 \centering
 \includegraphics[clip,trim=10mm 20mm 10mm 10mm, width=0.9\columnwidth]{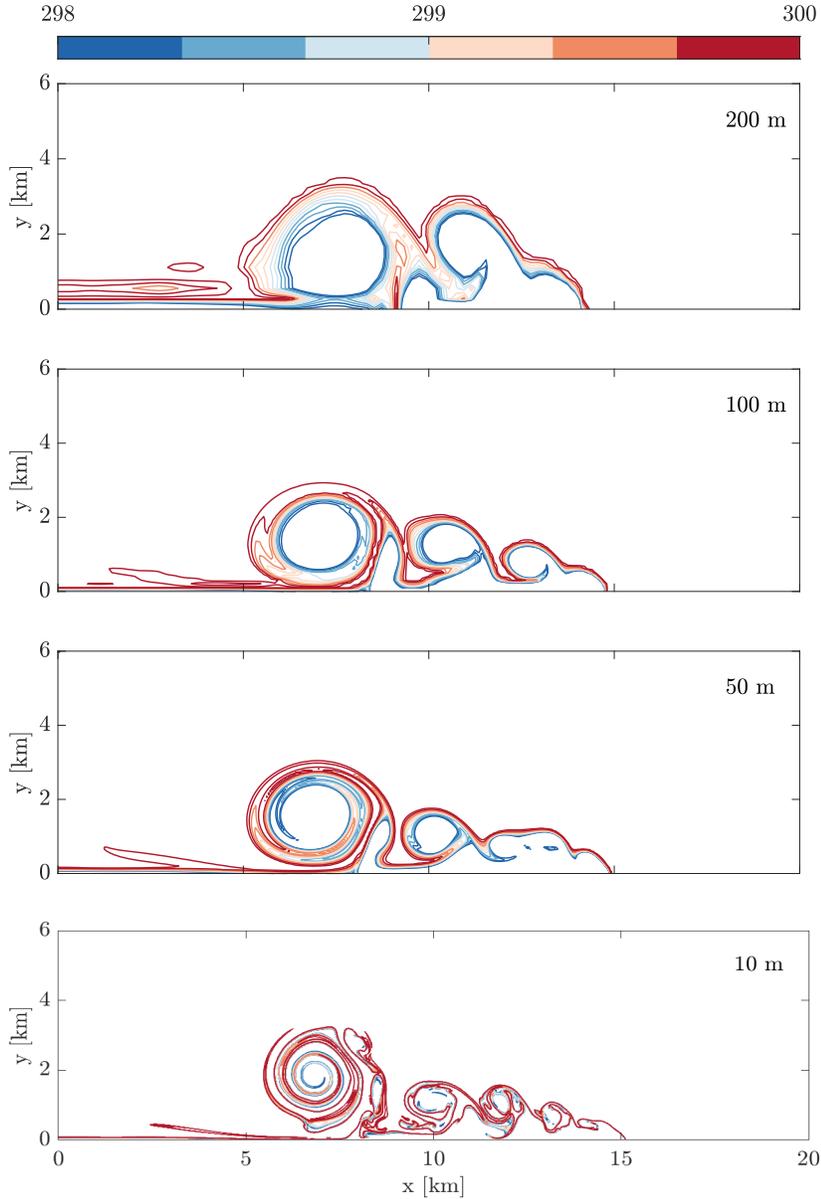}
  \caption{Contours of potential temperature [K] in the density current simulation at $900~\mathrm{s}$ at mesh resolutions of 200~m, 100~m, 50~m, and 10~m. Contours of potential temperature are shown at increments of  $0.2~\mathrm{K}$.}
  \label{fig:density_current}
\end{figure}

\Cref{fig:density_current} shows the potential temperature at $t = 900~\mathrm{s}$ for varying resolutions ranging from a homogeneous resolution of 200~m to 10~m. A uniform kinematic viscosity of $10~\mathrm{m^2~s^{-1}}$ is used to make the simulations comparable across the wide range of resolutions. Since the solutions are nearly horizontally symmetric about the center of the domain, only the right half of the bubble is shown. The flow exhibits Kelvin-Helmholtz instabilities that generate small scales. 

The numerical solutions exhibit increasingly detailed small-scale features as the resolution is increased. Even at the coarsest resolution (200~m), the large-scale flow features are preserved, and there are no signs of spurious small-scale oscillations associated with numerical dispersion errors. These results suggest the robustness of the numerical scheme in capturing sharp gradients and turbulence, even at coarser resolutions.

\subsection{Rising Bubble}

The second test case is a rising dry bubble. The bubble is initialized as a perturbation to a uniform potential temperature field, following \citeA{Bryan04a}, with a peak amplitude of $2~\mathrm{K}$. As in the first test case, we use periodic horizontal boundary conditions. The domain is $20~\mathrm{km}$ wide and $10~\mathrm{km}$ high.

\begin{figure}[!htb!]
 \centering
 \includegraphics[clip,trim=10mm 0mm 10mm 0mm,width=0.95\columnwidth]{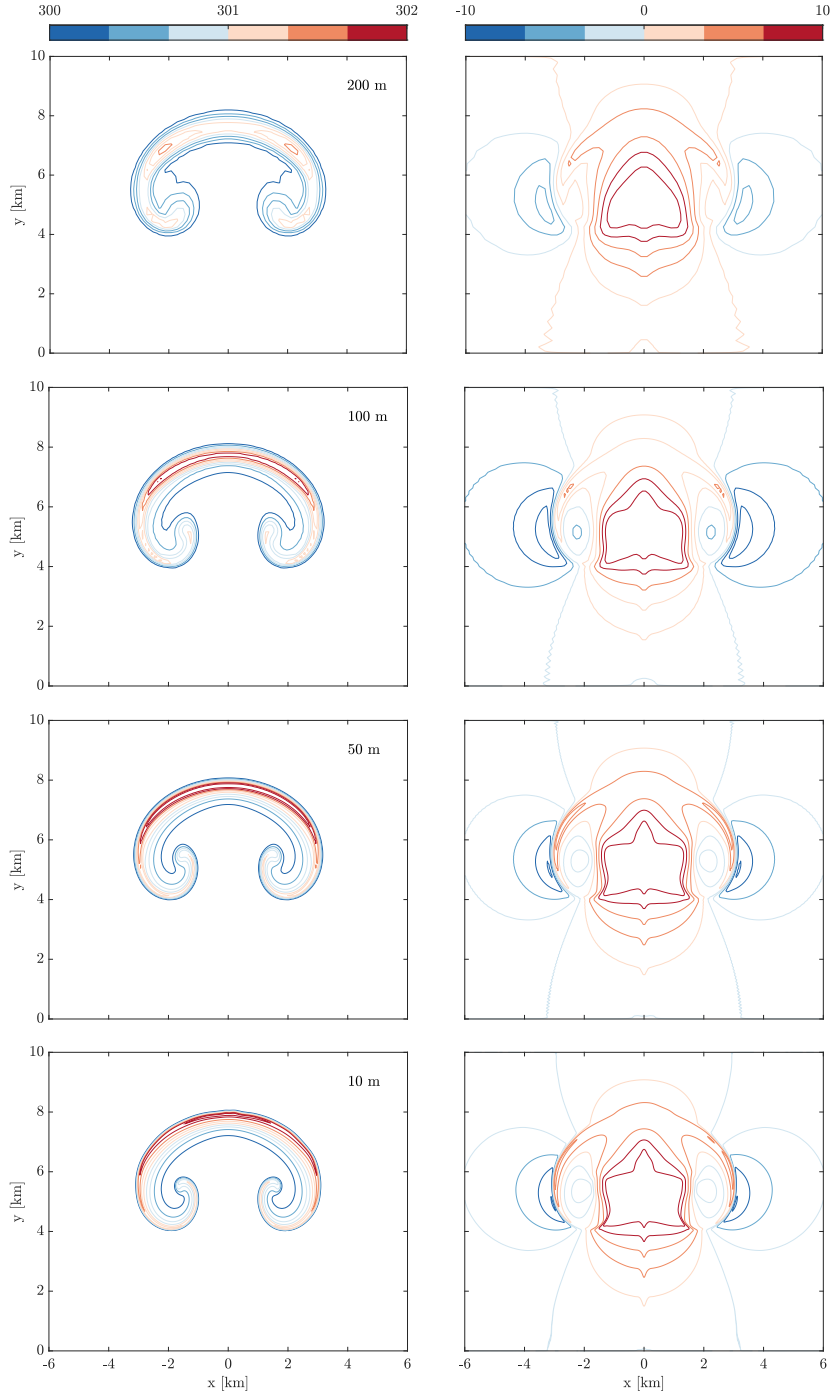}
  \caption{Contours of potential temperature [K] (left) and vertical velocity [$\mathrm{m~s^{-1}}$] (right) in the rising bubble simulation for mesh resolutions of $200$~m, $100$~m, $50$~m, and $10$~m. Contours of potential temperature and velocity are shown for increments of $0.2~\mathrm{K}$ and $2~\mathrm{m~s^{-1}}$, respectively.}
  \label{fig:bubble}
\end{figure}

\Cref{fig:bubble} shows the potential temperature at $t = 1000~\mathrm{s}$ for varying homogeneous resolutions ranging from $200~\mathrm{m}$ to $10~\mathrm{m}$. For this case, a uniform kinematic viscosity of $1~\mathrm{m^2~s^{-1}}$ was found to be adequate to ensure the simulations are comparable across the different resolutions.

As in the density current case, there are no spurious oscillations even for the simulations with coarser resolutions. The numerical solution is essentially converged at 50 m. The vertical velocity contours are nearly unchanged from the finest resolution down to 100 m resolution. This observed stability is due in great part to the QUICK scheme used in the scalar and momentum advection. Although the QUICK scheme achieves only a 2nd-order accurate approximation of the advective flux, the solutions seen in this case suggest a quality and fidelity of simulation comparable to that of the WENO schemes on staggered grids \cite{Pressel15a}.

\section{DYCOMS Simulation} \label{s:DYCOMS}

The first nocturnal research flight (RF01) of the Dynamics and Chemistry of Marine Stratocumulus (DYCOMS-II) field study \cite{Stevens03a} serves as the testbed of our low-cloud simulations. Among the attractive characteristics of this test case are the relative homogeneity of the environmental conditions, the absence of significant drizzle, and the persistence of a stable cloud layer. The basic state for RF01 is idealized as a quasi-two-layer structure in potential temperature $\theta_l$ and total-water specific humidity $q_t$ \cite{Stevens05a}. Forcings include geostrophic winds, large-scale subsidence, a simple parameterization of longwave radiation, and surface fluxes of latent and sensible heat.

We set the initial liquid-ice potential temperature and the initial total specific humidity in the mixed layer to be $\theta_{l} = 289~\mathrm{K}$ and $q_{t} = 9~\mathrm{g~kg^{-1}}$, respectively. This ensures that with our thermodynamics formulation and constants, we obtain a cloud layer between $600$ and $840~\mathrm{m}$. The vertical domain extends to $1.5~\mathrm{km}$, with a no-slip, zero-flux lid at the top. The horizontal domain in the default case covers an area of $(4~\mathrm{km})^2$, with periodic horizontal boundary conditions.

The default simulation runs for 4 simulated hours on a grid of $128 \times 128 \times 256$ points with a uniform horizontal grid spacing of $32~\mathrm{m}$ and a uniform vertical grid spacing of $6~\mathrm{m}$. Although a vertical resolution of $5~\mathrm{m}$ or less is often desirable to capture the sharp temperature gradient at the inversion above the cloud top without generating spurious entrainment \cite{Mellado16a,Pressel17a,Mellado18a}, our simulation did not change materially as we increased the vertical resolution to finer than $6~\mathrm{m}$. A physical time step of 0.3~s (Courant number 0.3) is used in the default configuration.

\begin{figure}[!htb!]
 \centering
 \includegraphics[clip,trim=0mm 218mm 0mm 0mm,width=\columnwidth]{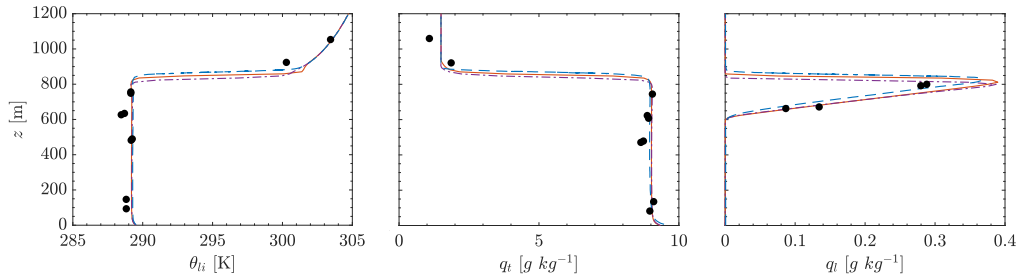}
  \caption{Profile of mean state specific humidity and temperature in DYCOMS as observed (points), from our simulation averaged over the 4th hour using the QUICK scheme (red solid lines) and the 3rd-order WENO scheme (violet dash-dotted lines), and from an implicit LES \protect\cite{Pressel17a} using a nominally 5th-order WENO scheme (blue dashed lines).}
  \label{fig:dycoms_mean_q_and_theta}
\end{figure}

\begin{figure}[!htb!]
 \centering
 \includegraphics[clip,trim=0mm 196mm 0mm 0mm,width=\columnwidth]{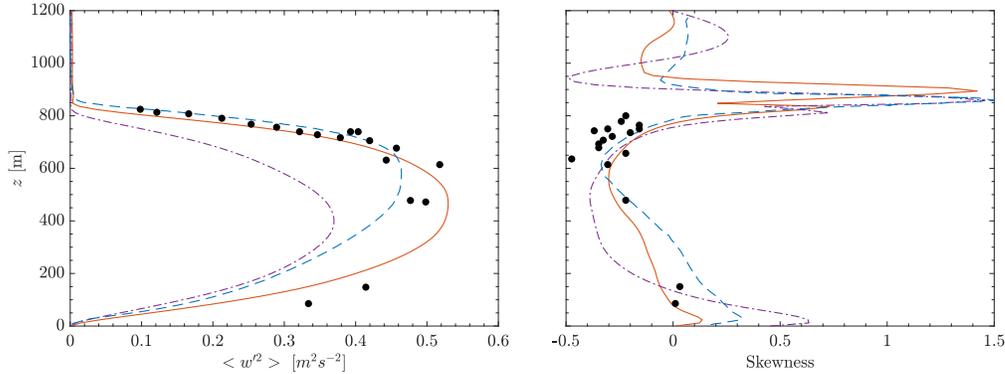}
  \caption{Profile of the variance and skewness of the vertical velocity in DYCOMS as observed (points), from our simulation averaged over the 4th hour using the QUICK scheme (red solid lines) and the 3rd-order WENO scheme (violet dash-dotted lines), and from an implicit LES \protect\cite{Pressel17a} using a nominally 5th-order WENO scheme (blue dashed lines).}
  \label{fig:dycoms_ww_www}
\end{figure}

\subsection{Fidelity of Simulation\label{subsec:dycoms_fidelity_analysis}}

The mean vertical profiles and vertical velocity statistics closely match the observations from the research flight. Both the liquid potential temperature profile and the total-water specific humidity profile maintain their two-layer structure, with a well-mixed boundary layer below a cloud top  (\cref{fig:dycoms_mean_q_and_theta}). For comparison, we also show the vertical profiles for a simulation with a 3rd-order WENO scheme with a Lax-Friedrichs numerical flux and for the simulation from \citeA{Pressel17a} with an implicit LES using a nominally 5th-order WENO scheme on a staggered grid, a configuration that has been shown to perform well for stratocumulus simulations; in fact, this configuration at the resolution we use here performs favorably relative to simulations with oscillatory numerical schemes for the momentum equation on much higher-resolution (meter-scale) isotropic grids \cite{Schneider19a,Matheou18a,Mellado18a}. (The WENO scheme uses a 5th-order stencil for the flux reconstruction but is strictly only of 2nd-order accuracy for nonlinear problems on a staggered grid \cite{Mishra21a}; hence, it is only nominally 5th order.) Comparison with the 3rd- and nominally 5th-order WENO simulations shows a spurious mixing layer above the inversion in our simulation with the QUICK scheme; this is likely the result of the QUICK scheme not being monotone, leading to spurious mixing of oscillations by the subgrid-scale diffusion scheme at the sharp gradients near the inversion \cite{Bretherton99a,Pressel17a}. Except for the differences near the inversion, however, the simulations with the QUICK scheme and with the 3rd-order WENO scheme are similar in their ability to reproduce the mean-state profiles. 

The turbulent structure in the boundary layer becomes evident in the variance and skewness profiles of the vertical velocity (\cref{fig:dycoms_ww_www}). The variance peaks near the cloud base, consistent with the research flight observations and turbulence generation by latent heat release at that altitude. The skewness reveals preferential directions of turbulent vertical velocities. For example, positive vertical velocity skewness near the bottom is consistent with the presence of significant heat fluxes at the sea surface, which drive convection. On the other hand, the negative skewness near the cloud top is consistent with the presence of downdrafts driven by radiative cooling. Like the variance, the skewness in our simulations is consistent with the research flight observations. By contrast, most LES in the DYCOMS intercomparison study \cite{Stevens05} are unable to capture the negative skewness near the cloud top, likely because of excessive spurious mixing across the inversion. The vertical velocity statistics are an indication that our numerics with the QUICK advection scheme avoid the excessive generation of spurious mixing across the inversion at the cloud top, which occurs in many other LES. The fidelity of the vertical velocity statistics to observations is similar to that obtained with WENO schemes \cite{Pressel17a}. However, the 3rd-order WENO scheme with the Lax-Friedrichs numerical flux underestimates the magnitude of the vertical velocity variance, likely as the result of the dissipative numerical flux.

Our simulation maintains more  liquid water in the cloud (\cref{fig:dycoms_mean_q_and_theta}) than most other LES in the DYCOMS intercomparison study \cite{Stevens05}. %
LES often have difficulties maintaining a cloud layer with sufficient liquid water because of spurious numerical mixing of dry air across the inversion at the cloud top \cite{Pressel17a}, which warms and dries the cloud layer.

These results indicate a high fidelity of our LES to the observed flow statistics. Our LES does not suffer from the shortcomings in many LES that lead to spurious turbulent entrainment at the inversion and a decoupling boundary layer; it performs similarly well as implicit LES with WENO schemes \cite{Pressel17a}. Therefore, it can adequately capture low clouds and enable the investigation of the feedbacks that make low clouds such an important regulator of the strength of greenhouse warming.

\subsection{Time to Solution and Scaling Analysis\label{subsec:scaling}}

The discretization schemes described above lend themselves to parallelization algorithms that are well suited for the TPU infrastructure. However, increased parallelism generally comes at the expense of greater communication between processors. As the communication overhead begins to dominate, the marginal benefit from increased parallelism diminishes. To assess the appropriateness of the TPU simulation framework for this class of problems, it is thus imperative to measure how well the simulation runtime scales with increased parallelism.

We examine the scalability of the solver using the DYCOMS case as a testbed. The method for doing so is to measure the mean turnaround time for a time step under different mesh configurations. For a fair comparison between different configurations, we keep the spatial resolution at $\Delta x = \Delta y = 35~\mathrm{m}$ and $\Delta z = 6~\mathrm{m}$ and the Courant number at approximately 0.3 for all cases. We find impressive scaling, notwithstanding that each simulation step involves the solution of an elliptic (globally nonlocal) problem for the dynamic pressure correction.

\subsubsection{Weak Scaling}

We demonstrate weak scalability by fixing the local grid size per processor and considering an ever-growing computational domain. We use $N_k$ to denote the global grid size along dimension $k$, $\widehat{N}_k$ to denote the local per processor subgrid size along dimension $k$, and $P_k$ to denote the number of processors assigned to dimension $k$ in the computational topology. For the first analysis, the computational domain per TPU core is fixed with a size of $\widehat{N}_x \times \widehat{N}_y \times \widehat{N}_z=1024 \times 36 \times 1024$ grid points, which is about the largest partition size that can fit in the TPU RAM considering the data requirements of this simulation. \Cref{tab:weak_scaling_large} shows that the turnaround time remains virtually unchanged as the number of TPU cores grows from $16$ to $2048$, corresponding to an increase in total number of grid points from 533M to 68.2B and in physical domain size from $36~\mathrm{km} \times 18~\mathrm{km}$ to $(286~\mathrm{km})^2$.

We repeat this weak scaling analysis using a smaller partition size that is more commonly encountered in atmospheric simulations. \Cref{tab:weak_scaling_small} demonstrates weak scalability when the computational domain per TPU core is fixed with a size of $\widehat{N}_x\times \widehat{N}_y\times \widehat{N}_z=128 \times 10 \times 256$. With about $10 \times$ speedup over real-time evolution (10 simulated days per day, SDPD) and a 35-meter horizontal resolution, the largest physical domain attainable with 2048 TPU cores in this configuration is $34.7~\mathrm{km} \times 53.8~\mathrm{km}$. (Since we are using only a quarter TPU pod, the largest domain size attainable on a full TPU pod, with 8192 cores, would be $69.4~\mathrm{km} \times 107.4~\mathrm{km}$.)

\begin{figure}[!htb!]
    \centering
    \includegraphics[width=0.6\textwidth]{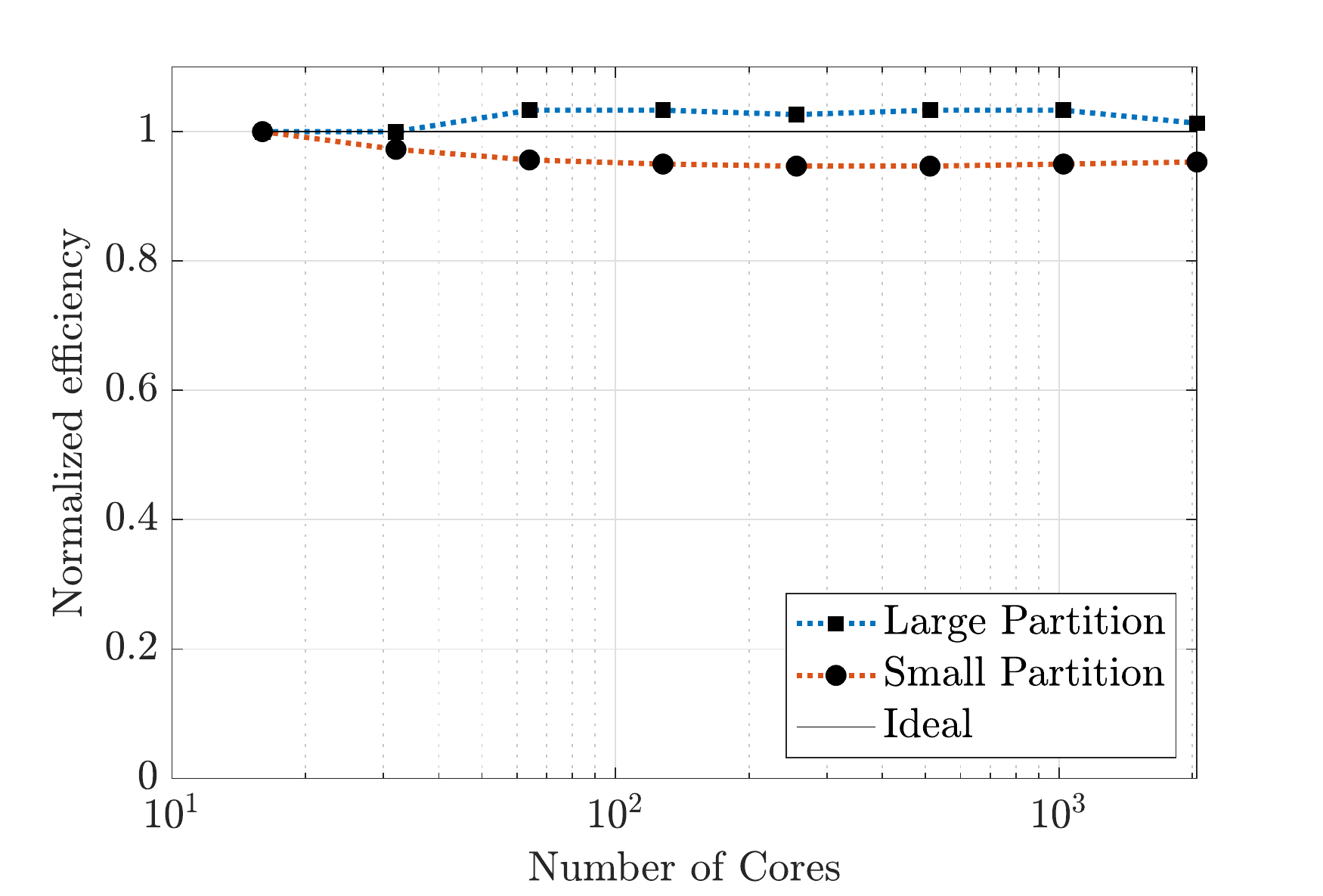}
    \caption{Normalized efficiency with respect to the weak scaling analysis.}
    \label{fig:weak_scaling}
\end{figure}

\Cref{fig:weak_scaling} shows the efficiency curves for these two cases, normalized relative to the smallest simulation. It is worth noting that in the weak scaling analysis with the large partitions, the two smallest simulations are in fact less efficient than the larger ones. This may seem surprising at first, as efficiency normally decreases with the problem size. However, this behavior is most likely a consequence of saturating the partition memory, which leads to variations in memory bandwidth utilization, which seem to penalize the performance of smaller TPU configurations more severely. This behavior is not seen in the second weak scaling analysis, which uses a significantly smaller partition size.

\subsubsection{Strong Scaling}

\begin{figure}[!htb!]
    \centering
    \includegraphics[width=0.6\textwidth]{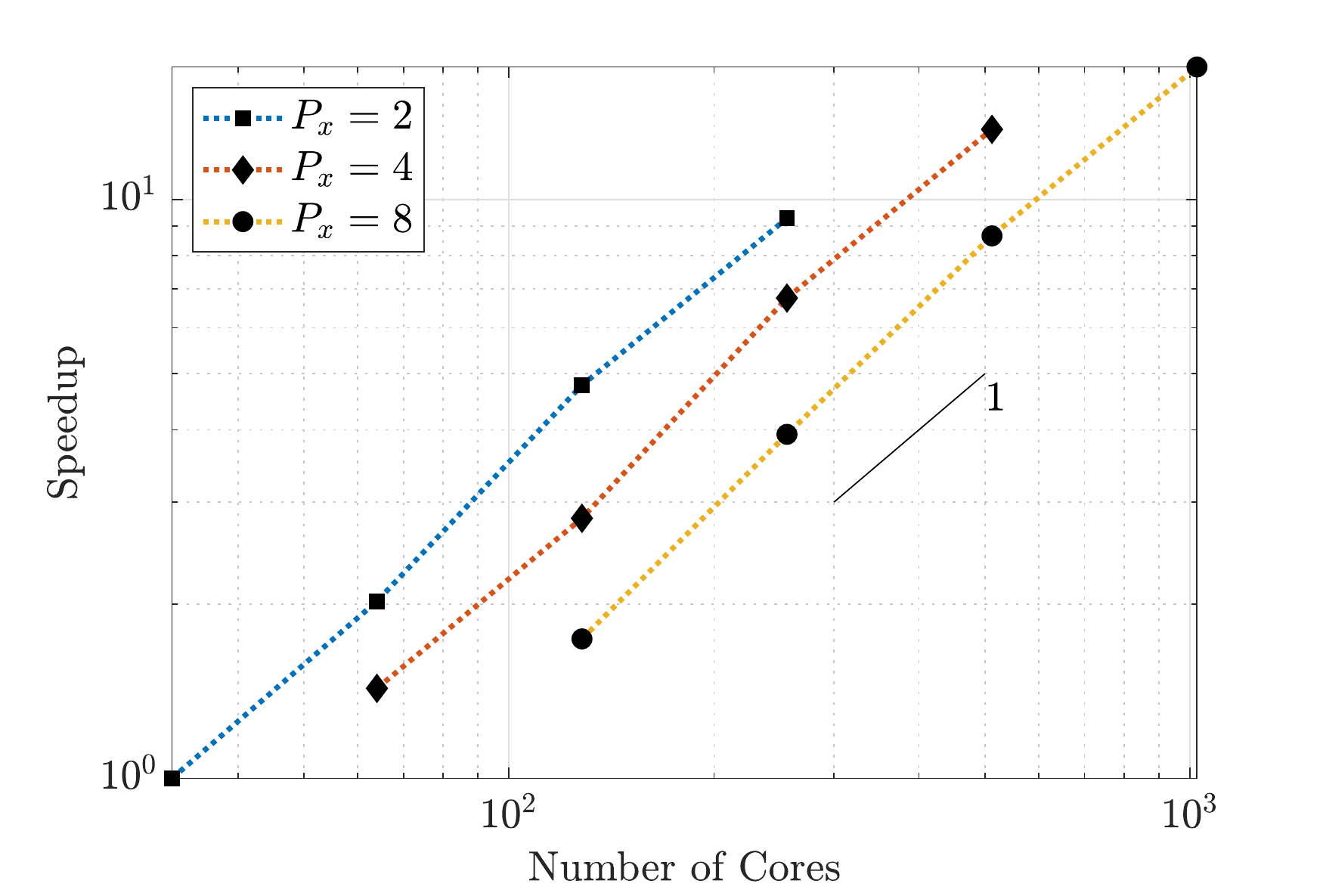}
    \caption{Strong scaling for different partitions (as in~\cref{tab:strong_scaling}).}
    \label{fig:strong_scaling}
\end{figure}

We now consider how the solver scales when we increase parallelism for a fixed global problem size. Throughout this analysis, the vertical dimension has only a single partition with a total of 128 levels. The total number of grid points is fixed at 134M. Three cases are considered: (i) 2 partitions in the $x$ direction, (ii) 4 partitions in the $x$ direction, and (iii) 8 partitions in the $x$ direction. In each of the three cases, we try multiple partitions in the $y$ direction, starting with 16 and scaling all the way up to 128 partitions. As seen in \cref{tab:strong_scaling}, the analysis consists of increasing parallelism while proportionately reducing the workload per processor. Each subsequent row doubles the number of cores assigned to the $y$ dimension while simultaneously halving the number of grid points per core along that dimension. The measured speedup relative to real time reaches a maximum of $14.08$ in the configuration with 1024 cores and the smallest partition size. The speedup is illustrated in \cref{fig:strong_scaling}. In all cases, the speedup curve shows clear evidence of linear (i.e., perfect) strong scaling.

\begin{table}[!htb!]
\centering
\footnotesize
\caption{Simulation configurations for weak scalability analysis using large partitions of size $1024 \times 36 \times 1024$ grid points per TPU core. The grid dimensions indicated in the middle columns do not include ghost points. The last column shows the simulated time relative to real time in simulated days per day (SDPD).}
\begin{tabular}{c c c c | c c c c | c}
\multicolumn{4}{c|}{Number of cores}&\multicolumn{4}{c|}{Grid size} & SDPD\\
$P_\text{tot}$ &$P_x$ & $P_y$ & $P_z$ &  $N_\text{tot}$ & $N_x\ (L_x)$ & $N_y\ (L_y)$ & $N_z\ (L_z)$ \\
     \hline
     16 & 1 & 16 & 1 & 533M & 1020 (35.7~{km}) & 512 (17.9~{km}) & 1020 (6.1~{km}) & 0.19 \\
     32 & 1 & 32 & 1 & 1.1B & 1020 (35.7~{km}) & 1024 (35.8~{km})& 1020 (6.1~{km}) & 0.19 \\
     64 & 2 & 32 & 1 & 2.1B & 2040 (71.4~{km}) & 1024 (35.8~{km})& 1020 (6.1~{km}) & 0.20 \\
     128 & 2 & 64 & 1 & 4.3B & 2040 (71.4~{km}) & 2048 (71.7~{km}) & 1020 (6.1~{km}) & 0.20 \\
     256 & 4 & 64 & 1 & 8.5B & 4080 (142.8~{km}) & 2048 (71.7~{km}) & 1020 (6.1~{km}) & 0.20 \\
     512 & 4 & 128 & 1 & 17.1B & 4080 (142.8~{km}) & 4096 (143.4~{km}) & 1020 (6.1~{km}) & 0.20 \\
     1024 & 8 & 128 & 1 & 34.1B & 8160 (285.6~{km}) & 4096 (143.4~{km}) & 1020 (6.1~{km}) & 0.20 \\
     2048 & 8 & 256 & 1 & 68.2B & 8160 (285.6~{km}) & 8192 (286.7~{km}) & 1020 (6.1~{km}) & 0.19 \\
\end{tabular}
\label{tab:weak_scaling_large}
\end{table}

\begin{table}[!htb!]
\centering
\footnotesize
\caption{Weak scalability analysis with a more typical partition of size $128 \times 10 \times 256$ per TPU core.}
\begin{tabular}{c c c c | c c c c | c }
\multicolumn{4}{c|}{Number of cores}&\multicolumn{4}{c|}{Grid size} & SDPD\\
$P_\text{tot}$ &$P_x$ & $P_y$ & $P_z$ &  $N_\text{tot}$ & $N_x\ (L_x)$ & $N_y\ (L_y)$ & $N_z\ (L_z)$ & \\
     \hline
     16 & 1 & 16 & 1 & 3M & 124 (4.3~{km}) & 96 (3.4~{km}) & 252 (1.5~{km}) & 10.56 \\
     32 & 1 & 32 & 1 & 6M & 124 (4.3~{km}) & 192 (6.7~{km})& 252 (1.5~{km}) & 10.27 \\
     64 & 2 & 32 & 1 & 12M & 248 (8.7~{km}) & 192 (6.7~{km})& 252 (1.5~{km}) & 10.10 \\
     128 & 2 & 64 & 1 & 24M & 248 (8.7~{km}) & 384 (13.4~{km}) & 252 (1.5~{km}) & 10.03 \\
     256 & 4 & 64 & 1 & 48M & 496 (17.4~{km}) & 384 (13.4~{km}) & 252 (1.5~{km}) & 10.00 \\
     512 & 4 & 128 & 1 & 96M & 496 (17.4~{km}) & 768 (26.9~{km}) & 252 (1.5~{km}) & 10.00 \\
     1024 & 8 & 128 & 1 & 192M & 992 (34.7~{km}) & 768 (26.9~{km}) & 252 (1.5~{km}) & 10.03 \\
     2048 & 8 & 256 & 1 & 384M & 992 (34.7~{km}) & 1536 (53.8~{km}) & 252 (1.5~{km}) & 10.07 \\
\end{tabular}
\label{tab:weak_scaling_small}
\end{table}

\begin{table}[!htb!]
\centering
\footnotesize
\caption{Partitions for strong scalability analysis ($N_\text{tot}=134\mathrm{M}$).}
\begin{tabular}{ c c c c | c c c | c }
\multicolumn{4}{c|}{Number of cores}&\multicolumn{3}{c|}{Partition size} & SDPD \\
$P_\text{tot}$ &$P_x$ & $P_y$ & $P_z$ &  $\widehat{N}_x$ & $\widehat{N}_y$ & $\widehat{N}_z$ &  \\
     \hline
     32 & 2 & 16 & 1 & 512 & 64 & 128 & 0.83 \\
     64 & 2 & 32 & 1 & 512 & 32 & 128 & 1.68 \\
     128 & 2 & 64 & 1 & 512 & 16 & 128 & 3.97 \\
     256 & 2 & 128 & 1 & 512 & 8 & 128 & 7.71 \\
     \hline
     64 & 4 & 16 & 1 & 256 & 64 & 128 & 1.19 \\
     128 & 4 & 32 & 1 & 256 & 32 & 128 & 2.34 \\
     256 & 4 & 64 & 1 & 256 & 16 & 128 & 5.62 \\
     512 & 4 & 128 & 1 & 256 & 8 & 128 & 10.99 \\
     \hline
     128 & 8 & 16 & 1 & 128 & 64 & 128 & 1.45 \\
     256 & 8 & 32 & 1 & 128 & 32 & 128 & 3.27 \\
     512 & 8 & 64 & 1 & 128 & 16 & 128 & 7.19 \\
     1024 & 8 & 128 & 1 & 128 & 8 & 128 & 14.08 \\
\end{tabular}
\label{tab:strong_scaling}
\end{table}

\subsection{Taking LES of Clouds to the Macroscale}

\begin{figure}[!htb!]
  \centering
  \includegraphics[clip,trim=0mm 70mm 0mm 0mm,width=\columnwidth]{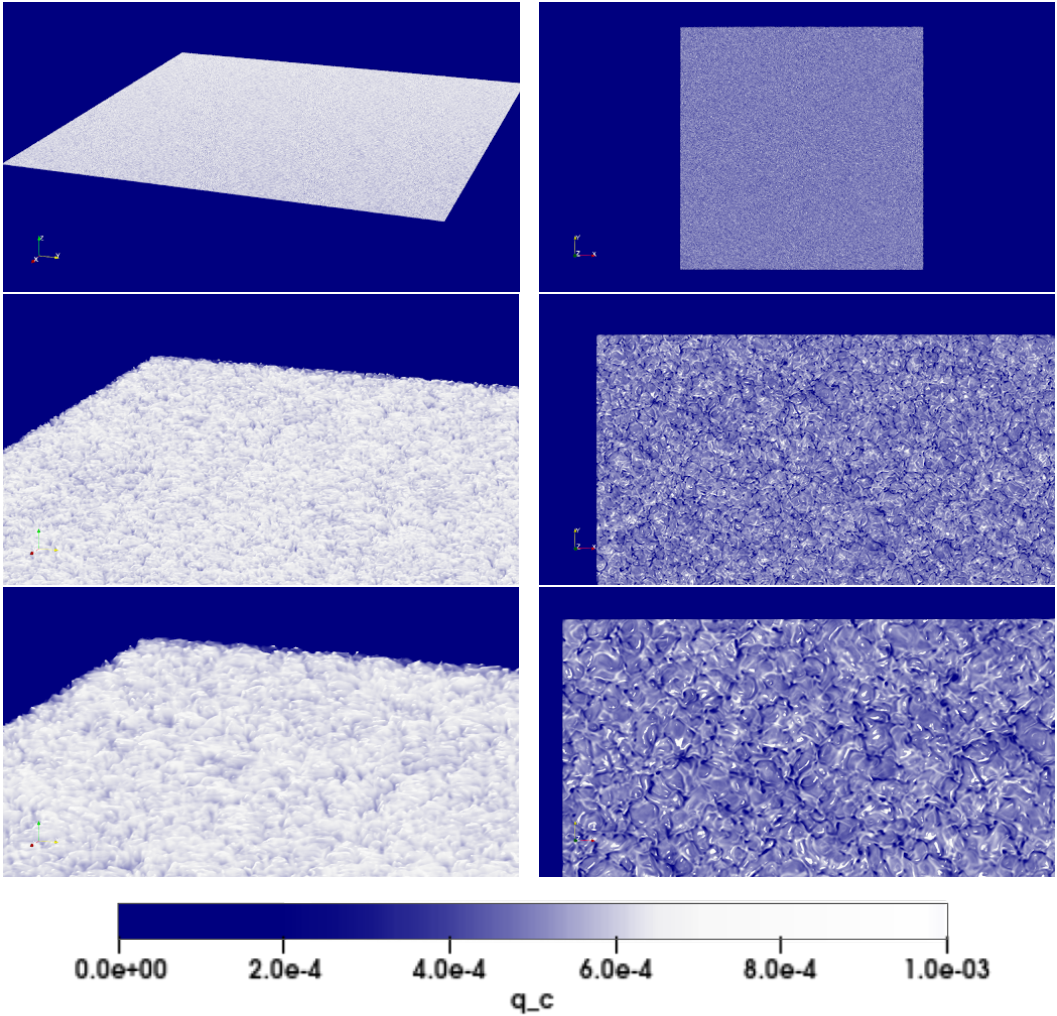}
\caption{Volume rendering of the instantaneous cloud water specific humidity $q_c$ of a simulated stratocumulus cloud covering a horizontal $(285~\mathrm{km})^2$ footprint after 4 simulated hours. A modified DYCOMS RF01 configuration with a slightly different initial liquid-ice potential temperature was used. (left column) Oblique view and (right column) normal view. (top) Entire domain; (middle) closeup of a corner: (oblique) $26~\mathrm{km}\times 26~\mathrm{km}$ and (normal) $52~\mathrm{km}\times 26~\mathrm{km}$; (bottom) further closeup of the same corner: (oblique) $13~\mathrm{km}\times13~\mathrm{km}$, and (normal)  $26~\mathrm{km}\times13~\mathrm{km}$.}
\label{fig:dycoms_large_scale}
\end{figure}

To demonstrate the TPU framework’s capabilities for simulating clouds on large scales, we simulated 4 hours of the DYCOMS conditions on a domain of size $285~\mathrm{km} \times 285~\mathrm{km} \times 2~\mathrm{km}$ using a typical DYCOMS resolution of $35~\mathrm{m} \times 6~\mathrm{m}$. The simulation runs on a mesh of 32 billion grid points and requires a little less than 20 wallclock hours to simulate 4 hours on 1024 TPU cores (70.4 petaflops at single precision). It should be noted that this simulation used a modified DYCOMS RF01 configuration with a slightly different initial liquid-ice potential temperature, which leads to a thicker stratocumulus cloud. The cloud layer is visualized in \cref{fig:dycoms_large_scale} at different scales. The visualization reveals large spatial variability of cloud water fraction, with occasional open cells.

These simulations demonstrate that low-cloud resolving LES are possible in domains the size of a grid box in a typical coarse-resolution climate model, which has a footprint of around $(100~\mathrm{km})^2$. It enables three-dimensional LES to be embedded in climate model grid boxes, to provide high-fidelity representations of cloud dynamics locally in them.

\section{Discussion and Conclusions} \label{s:concl}

We have demonstrated that LES of low clouds are possible on TPUs and achieve unprecedented weak and strong scaling at high numerical fidelity. Our LES code with a QUICK advection scheme for momentum and tracers demonstrates a fidelity to aircraft observations that is comparable with that obtained with WENO schemes at the same resolution, exceeding the fidelities achievable with oscillatory, numerical schemes, or combinations of oscillatory schemes for momentum and non-oscillatory schemes for tracers \cite{Pressel17a,Schneider19a}. At the meter-scale resolutions needed for resolving the computationally challenging stratocumulus clouds, we have shown that the code scales strongly and weakly up to 1024 and 2048 TPU cores, respectively, corresponding to a computational throughput of 70.4 and 140.8 petaflops. This opens up the large compute resources with fast chip-to-chip interconnects available on TPUs for low-cloud LES. For example, it means that LES with horizontal resolutions around $30~\mathrm{m}$ and vertical resolutions around $5~\mathrm{m}$ are achievable at 10 simulated days per wallclock day in domains the size of what is becoming a typical climate model grid column ($25$--$50~\mathrm{km}$ wide). Thus, it is possible to generate LES with an outer horizontal scale that is the same as the inner horizontal scale of climate models \cite{Schneider17a}.  

Our LES code and the compute resources available on TPUs enable the generation of large libraries of low-cloud simulations \cite{Shen22j}. These can be used both for quantitatively studying mechanisms underlying low-cloud feedbacks to climate change \cite{Bretherton15a} and as training data for parameterizations of low clouds for coarse-resolution climate models \cite{Couvreux21a,Hourdin21b,Lopez-Gomez22a}. The LES code described here is publicly available for this and similar purposes.

\section{Open Research}

The source code for all simulations described in this paper and used to produce the data displayed in the figures and tables is available at https://doi.org/10.5281/zenodo.7569544 \cite{Wang_2023_7569544} under the Apache License, Version 2.0.

\acknowledgments
We thank Jason Hickey for his guidance in the early stages of this research and Tianjian Lu for his valuable feedback. We also thank the reviewer for their insightful comments and suggestions.

\appendix
\section{Numerically Consistent Poisson Equation on Collocated Grids\label{sec:appendix_a}}
To eliminate the discrepancy between the numerical representation of the gradient and Laplacian operators in the Poisson~\cref{eq:pressure_correction} (which has been shown to have a dissipative effect on kinetic energy \cite{HAM_ETAL_ARB2004}), and to introduce coupling between nodes with odd and even indices, we add an additional correction term that takes the form of a fourth-order difference of the pressure correction $\delta p$ on the right-hand side of~\cref{eq:pressure_correction}. Specifically, applying the discrete divergence operator to~\cref{eq:momentum_correction} with the enforcement of mass conservation at sub-iteration $k+1$, we have
\begin{equation}
    \nabla\cdot\nabla(\alpha_0\delta p) = \frac{\alpha_0}{\Delta t}\nabla\cdot(\widehat{\rho_0\vec{u}}).
    \label{eq:div_momentum_eq_k+1}
\end{equation}
Subtracting~\cref{eq:div_momentum_eq_k+1} from~\cref{eq:pressure_correction} results in a correction term that takes the form:
\begin{equation}
    \mathcal{C}=(\nabla^2 - \nabla\cdot\nabla)(\alpha_0 \delta p).
    \label{eq:poisson_eq_correction}
\end{equation}
In a discrete representation in which the divergence operator is expressed by the 2nd-order central difference scheme,
\begin{equation}
    \nabla(\cdot)=\frac{(\cdot)_{l+1}-(\cdot)_{l-1}}{2\Delta_l},
\end{equation}
and the Laplacian operator is expressed as
\begin{equation}
    \nabla^2(\cdot)=\frac{(\cdot)_{l+1}-2(\cdot)_{l}+(\cdot)_{l-1}}{\Delta_l^2},
\end{equation}
the correction term in~\cref{eq:poisson_eq_correction} is computed numerically as
\begin{align}
    \mathcal{C} &= \frac{1}{2\Delta_l}[\nabla(\alpha_0\delta p))_{l+1}-(\nabla(\alpha_0\delta p))_{l-1}] - \frac{1}{\Delta_l^2}[\alpha_0\delta p)_{l+1}-2(\alpha_0\delta p)_{l}+(\alpha_0\delta p)_{l-1}]  \nonumber \\
    &= \frac{1}{4\Delta_l^2}[(\alpha_0\delta p)_{l+2}-2(\alpha_0\delta p)_{l}+(\alpha_0\delta p)_{l-2}] - \frac{1}{\Delta_l^2}[(\alpha_0\delta p)_{l+1}-2(\alpha_0\delta p)_{l}+(\alpha_0\delta p)_{l-1}]   \nonumber \\
    &= \frac{1}{4\Delta_l^2}[(\alpha_0\delta p)_{l+2}-4(\alpha_0\delta p)_{l+1}+6(\alpha_0\delta p)_l-4(\alpha_0\delta p)_{l-1}+(\alpha_0\delta p)_{l-2}].
    \label{eq:poisson_eq_correction_discrete}
\end{align}
To ensure~\cref{eq:pressure_correction} is solved with numerical consistency,~\cref{eq:poisson_eq_correction_discrete} is added to the divergence of the momentum on the right-hand side of the equation, which is:
\begin{equation}
    \nabla^2(\alpha_0\delta p)^{k+1}=\frac{\alpha_0}{\Delta t}\nabla\cdot(\widehat{\rho_0 \vec{u}})-\mathcal{C}^{k}.
\end{equation}
This ensures a numerically consistent treatment of the derivative operators and leads to coupling of nodes with even and odd indices.

It is worth noting that the explicit correction to the right-hand side of~\cref{eq:pressure_correction} derived in~\cref{eq:poisson_eq_correction_discrete} can be equivalently expressed as a correction to the cell-face momentum. As such it resembles an existing approach that achieves a 2nd-order accurate scheme on a collocated grid by introducing a correction to the cell-face momentum that is proportional to the pressure gradient \cite{MOIN_ETAL_JCP1998}. Despite the resemblance, there are fundamental differences between that approach and the one taken here. The correction in~\cref{eq:poisson_eq_correction_discrete} is a function of the previous sub-iteration pressure correction $\delta p^k$ and not the actual pressure field. Additionally, the functional form of the two corrections are markedly different: our cell-face momentum correction is proportional to the 3rd-order difference of $\delta p^k$, whereas the correction from \citeA{MOIN_ETAL_JCP1998} is proportional to the first-order difference of the pressure field. From the standpoint of numerical stability, the former has the advantage that the third derivative will be large if the pressure oscillates rapidly, which will trigger the correction and result in a smoother pressure field.

\bibliography{lit,cloud-sim}

\end{document}